\documentstyle[epsfig,sprocl]{article}

\include{epsfig}

\def\Journal#1#2#3#4{{#1} {\bf #2}, #3 (#4)}

\def\NPB{{\em Nucl. Phys.} B}
\def\NPA{{\em Nucl. Phys.} A}

\def\PRL{\em Phys. Rev. Lett.}

\def\PRE{{\em Phys. Rev.} E}
\def\PRP{\em Phys. Rep.}

\def\be{\begin{equation}}
\def\ee{\end{equation}}
\def\bea{\begin{eqnarray}}
\def\eea{\end{eqnarray}}

\begin{document}

\begin{flushright}
UA-NPPS 01/12\\
November 2001
\end{flushright}
\vspace{1.0cm}

\title{CRITICAL FLUCTUATIONS AT RHIC\footnote{Presented at the XXXI International
Symposium on Multiparticle Dynamics, September 1-7, 2001, Datong, China}}

\author{N. G. ANTONIOU, F. K. DIAKONOS and A. S. KAPOYANNIS}

\address{Department of Physics, University of Athens, GR-15771,\\
Athens, Greece}

\maketitle\abstracts{On the basis of universal scaling properties, we claim that in $Au+Au$ collisions at RHIC,
the QCD critical point is within reach. The signal turns out to be an extended plateau of net baryons
in rapidity with approximate height $n_b \approx 15$ and a strong intermittency pattern
with index $s_2=1/6$ in rapidity fluctuations. A window also exists, to reach the critical
point at the SPS, especially in $Si+Si$ collisions at maximal energy.}

\section{Introduction}

The existence of a critical point in the phase diagram of QCD is a fundamental property
of strong interactions and therefore its experimental verification is of great importance.
It is the remnant of a tricritical point, associated with the chiral phase transition, 
located at the end of a critical line of first order in the phase diagram \cite{Wil00}.
Theoretical efforts to reveal the existence and location of the critical point, from first
principles, are now in progress, in studies based on the Lee-Yang theory of phase 
transitions, treated on the lattice \cite{Fod01}. On the other hand, the phenomenology
of the QCD critical point in heavy-ion physics \cite{Wil00,Ant01} becomes more and more 
promising due to the accumulation of measurements both from SPS and recently from RHIC.
In this presentation we explore the scaling properties associated with the universality
class of the QCD critical point in order (a) to derive quantitative criteria for a given
experiment to drive the system close to the critical point and (b) to construct a theory
of critical fluctuations in the net-baryon sector which can be tested in these experiments.
On the basis of our predictions, a set of necessary and sufficient conditions is proposed,
as a sharp signature of the QCD critical point at RHIC.

\section{Net-baryon scaling}

The natural order parameter, incorporating the universal behaviour of quark matter near the
critical point, is given by the classical behaviour of an isoscalar field ($\sigma$-field)
in $3d$ with zero mass at the critical point. The fluctuations of the $\sigma$-field 
manifest themselves as density fluctuations in the pion sector and can be detected in 
heavy-ion experiments, especially in low multiplicity events \cite{Ant01}. Fluctuations
of the same origin are expected to appear also in the net-baryon sector since near the
critical point $(\rho_c,T_c)$ an equivalent order parameter is given by the difference
of the net-baryon density $(\rho)$ from its critical value, $\langle \bar{q} q \rangle
\sim (\rho - \rho_c)$. This new order parameter obeys the same scaling laws as the 
$sigma$-field, dictated by the critical exponents of the $3d$ Ising system $(\beta
\approx \frac{1}{3}, \delta \approx 5, \nu \approx \frac{2}{3}, \eta \approx 0)$.
Approaching the critical point $(T \to T_c)$, the order parameter $m(\vec{x})=\rho-\rho_c$
satisfies, along the rapidity axis, a scaling law of the general form \cite{Anto01}:
\begin{eqnarray}
m(y)&\approx& t^{\beta} [ F_o(y/L) + t F_1(y/L)]~~; \nonumber\\
m(y)&=&A_{\perp}^{-2/3} n_b(y)-\rho_c
\label{eq:eq1}
\end{eqnarray}
where $t \equiv \frac{T_c - T}{T_c}$ $(T \leq T_c)$ and terms of order $O(t^{\beta+2})$
have been neglected. Eq.(\ref{eq:eq1}) describes the critical behaviour of the net-baryon fluid, confined along the rapidity axis in the region $0 \leq y \leq L$ and extended in a
large area of radius $R_{\perp} \sim A_{\perp}^{1/3}$ in the transverse space. The order
parameter $m(y)$ is associated with the net-baryon rapidity density $n_b(y)$ produced in
a heavy-ion collision $(A+A)$. If $A_t$ is the total number of participants in such a 
collision, equation (\ref{eq:eq1}) leads to the following scaling law in the neighbourhood 
of the critical point, along the freeze-out line of the process \cite{Anto01}:
\begin{eqnarray}
A_{\perp}^{-2/3} n_b(0)&=&\rho_c + \frac{4^{\beta/\nu}}
{B(1-\frac{\beta}{\nu},1-\frac{\beta}{\nu})} [f(z_c,\rho_c)]^{\beta}
+ C [f(z_c,\rho_c)]^{\beta+1}~~~~;~~~~z_c \geq \rho_c \nonumber \\
A_{\perp}^{-2/3} n_b(0)&\approx&z_c~~~~;~~~~z_c < \rho_c
\label{eq:eq2}
\end{eqnarray}
The scaling variable in eq.(\ref{eq:eq2}) is $z_c=A_{\perp}^{-2/3} A_t L^{-1}$ 
$(A_{\perp} \approx \frac{A_t}{2})$ and the scaling function $f(z_c,\rho_c)$ vanishes for
$z_c=\rho_c$: $f(z_c,\rho_c)=\frac{1}{G}(-1+\sqrt{1+2G(z_c-\rho_c)^{1/\beta}})$. The 
constants $C$ and $G$ give a measure of the non leading term $F_1$ in eq.(\ref{eq:eq1})
($C \sim F(1/2)$, $G \sim \int_0^1 F_1(x) dx$) and they are fixed by the measurements of 
net baryons at the SPS. In fact the trend of the data suggests the existence of a critical
point $(\rho_c \approx 0.8)$ near the chemical potential of $S+S$ central collisions 
(Fig.~1).

\begin{figure}[htb]
\begin{center}
\mbox{\epsfig{file=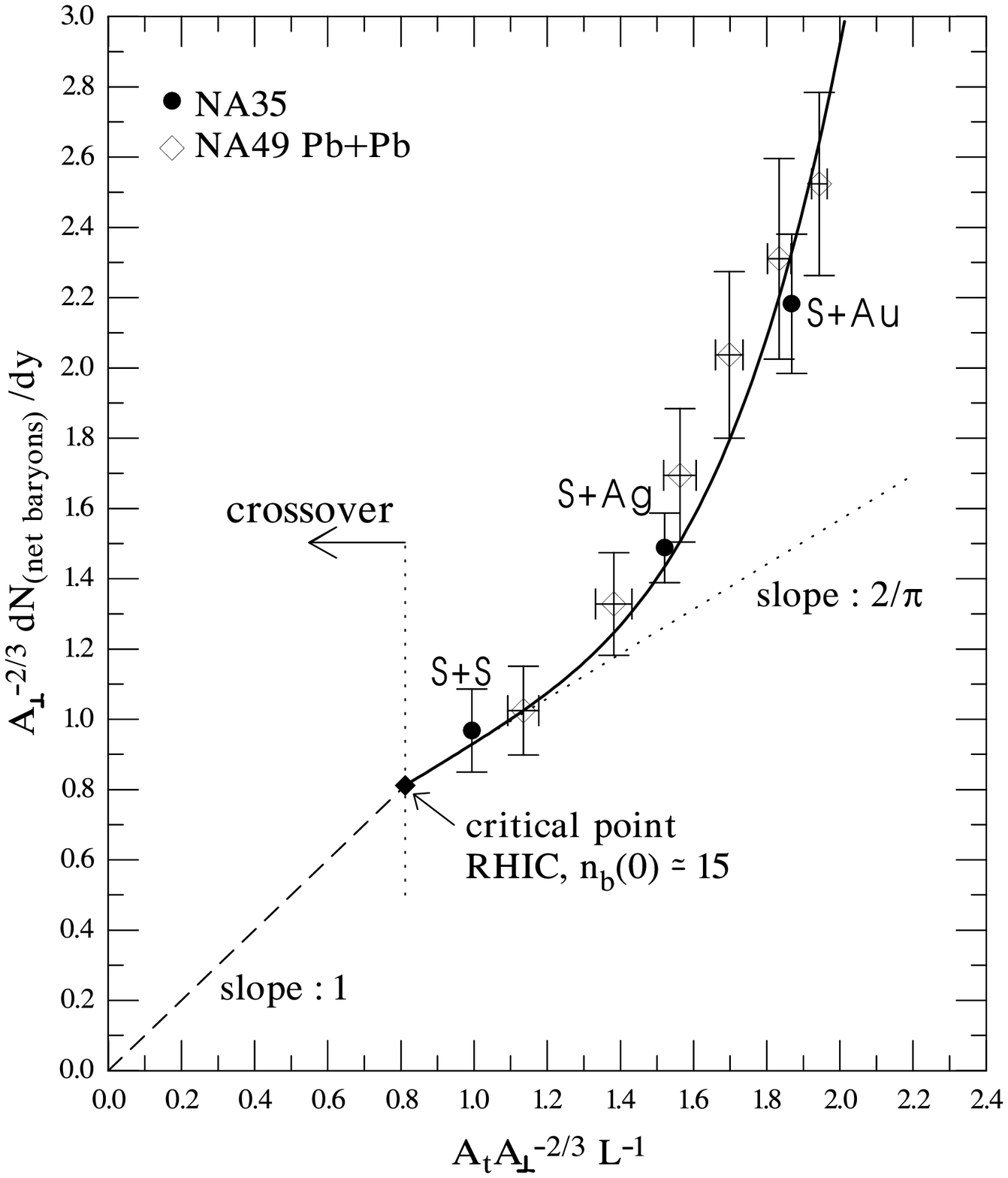,width=0.9\textwidth,angle=0}}
\caption{The diagram showing the proximity of heavy-ion experiments to
the critical point. At RHIC, the critical point corresponds to the values
$n_b \approx 15$, $A_t \approx 165$.}
\label{fig:1}
\end{center}
\end{figure}

The condition for a given experiment to reach the critical point, turns out to
be the development of an extended plateau of net baryons in rapidity with height
$n_b \approx 0.13 L^2$. At RHIC, this condition gives $n_b \approx 15$ and 
$A_t \approx 165$ whereas at the SPS the corresponding values are smaller, $n_b \approx 5$
and $A_t \approx 30$. As a result, the critical point may become accessible at RHIC with
$Au+Au$ collisions whereas at the SPS it can only be reached in collisions of medium-size
nuclei ($S+S$, $Si+Si$).

\section{Critical fluctuations}

The net-baryon system, produced in a particular class of experiments with heavy ions, 
develops strong critical fluctuations in rapidity density if the constraints, discussed in
the previous section, are satisfied. The origin of these fluctuations can be traced in the
free energy of the $\sigma$-field at $T=T_c$, given by the effective action \cite{Ant01}:
\begin{equation}
\Gamma_c=\frac{\pi R_{\perp}^2 \beta_c}{\tau_c} \int_{\delta y} dy [\frac{1}{2}
\left(\frac{\partial \sigma}{\partial y}\right)^2  
+ 2 \tau_c^2 \beta_c^{-4}
\left(\beta_c \sigma\right)^{\delta +1}]~~~~;~~~~\beta_c \equiv T_c^{-1}
\label{eq:eq3}
\end{equation}
where $\delta$ is the appropriate critical exponent at $T=T_c$ $(\delta \approx 5)$ and
$\delta y$ the size of a critical cluster. The manifestation of $\sigma$-fluctuations in
phase space, at the level of multiparticle production, is expected both in the pion sector
$(\sigma \to \pi^+ \pi^-)$ and the net-baryon sector. At RHIC, the detection of 
fluctuations in the pion sector is not easy because the multiplicity in central collisions
is high and the excess of pions, coming from sigmas, forms a subsystem of rather low
multiplicity $(n^{(\sigma)}_{ch} \approx 150)$ in each event \cite{Ant01}. Therefore the
measurement of net-baryon fluctuations becomes a superior proposal for RHIC, in the search
for the QCD critical point. In order to reveal the nature of the critical fluctuations in
the net-baryon sector, we introduce the order parameter $m(y)$ through the equations:
\begin{eqnarray}
\sigma(y)\approx F \beta_c^2 m(y)~~;~~F \equiv -\frac{\lambda \langle
\bar{q}q \rangle_o}{\rho_c}~~;~~\langle \bar{q} q \rangle_o
\approx -3~fm^{-3} \nonumber \\
\Gamma_c \approx g_1 \int_{\delta y} dy \left[ \frac{1}{2} \left(
\frac{\partial \hat{m}}{\partial y} \right)^2 + g_2 \vert \hat{m}
\vert^{\delta +1} \right]~~;~~\hat{m}(y)=\beta_c^3 m(y)
\label{eq:eq4}
\end{eqnarray}
where $g_1 \equiv F^2 \left(\frac{\pi R_{\perp}}{\tau_c \beta_c}\right)$, $g_2 \equiv
2 F^4 \left( \frac{\tau_c}{\beta_c}\right)^2$. The partition function 
$Z=\int {\cal{D}}[\hat{m}] e^{-\Gamma_c[\hat{m}]}$ for each cluster is saturated by
instanton-like configurations \cite{Ant98} which for $\delta y \leq \delta_c$ lead to
self-similar structures characterized by a pair-correlation of the form:
\begin{equation}
\langle \hat{m}(y) \hat{m}(0) \rangle \approx \frac{5}{6}
\frac{\Gamma(1/3)}{\Gamma(1/6)} \left(\frac{\pi R_{\perp}^2 \tau_c}{\beta_c^3}
\right) F^{-1} y ^{-\frac{1}{\delta +1}}
\label{eq:eq5}
\end{equation}
The maximal size of these fractal clusters is 
$\delta_c \approx \left(\frac{\pi R_{\perp}^2}{16 \tau_c^2}\right)^{2/3}$, 
according to the geometrical description of a critical system \cite{Ant98}. The
dimensionless parameter $F$ is of the order $10^2$ and the size $\delta_c$, on general
grounds $(R_{\perp} \stackrel{<}{~} 2 \tau_c)$, is of the order of one 
($\delta_c \approx 0.35$). As a result the global baryonic system (at RHIC the size of the
system is $L \approx 11$) develops fluctuations at all scales in rapidity since the direct
correlation (\ref{eq:eq5}) propagates along the entire system through the cooperation
of many self-similar clusters of relatively small size. We have quantified this mechanism
in a Monte-Carlo simulation for the conditions at RHIC, associated with the critical point
($L \approx 11$, $A_t \approx 165$). In Fig.~2 a typical event of net baryons at RHIC is
presented, together with the corresponding intermittency pattern \cite{Bia86} revealing the
nature of the critical fluctuations in rapidity. The intermittency exponent of the second
moment $F_2$ $(s_2 \approx 0.18)$ is close to the value $d-d_F = 1/6$ ($d=1$,
$d_F=\frac{\delta}{\delta +1}$) expected from the universality class of the critical point. It is
of interest to note that a stronger intermittency effect is expected in two and three
dimensional events.

\begin{figure}[htb]
\begin{center}
\mbox{\epsfig{file=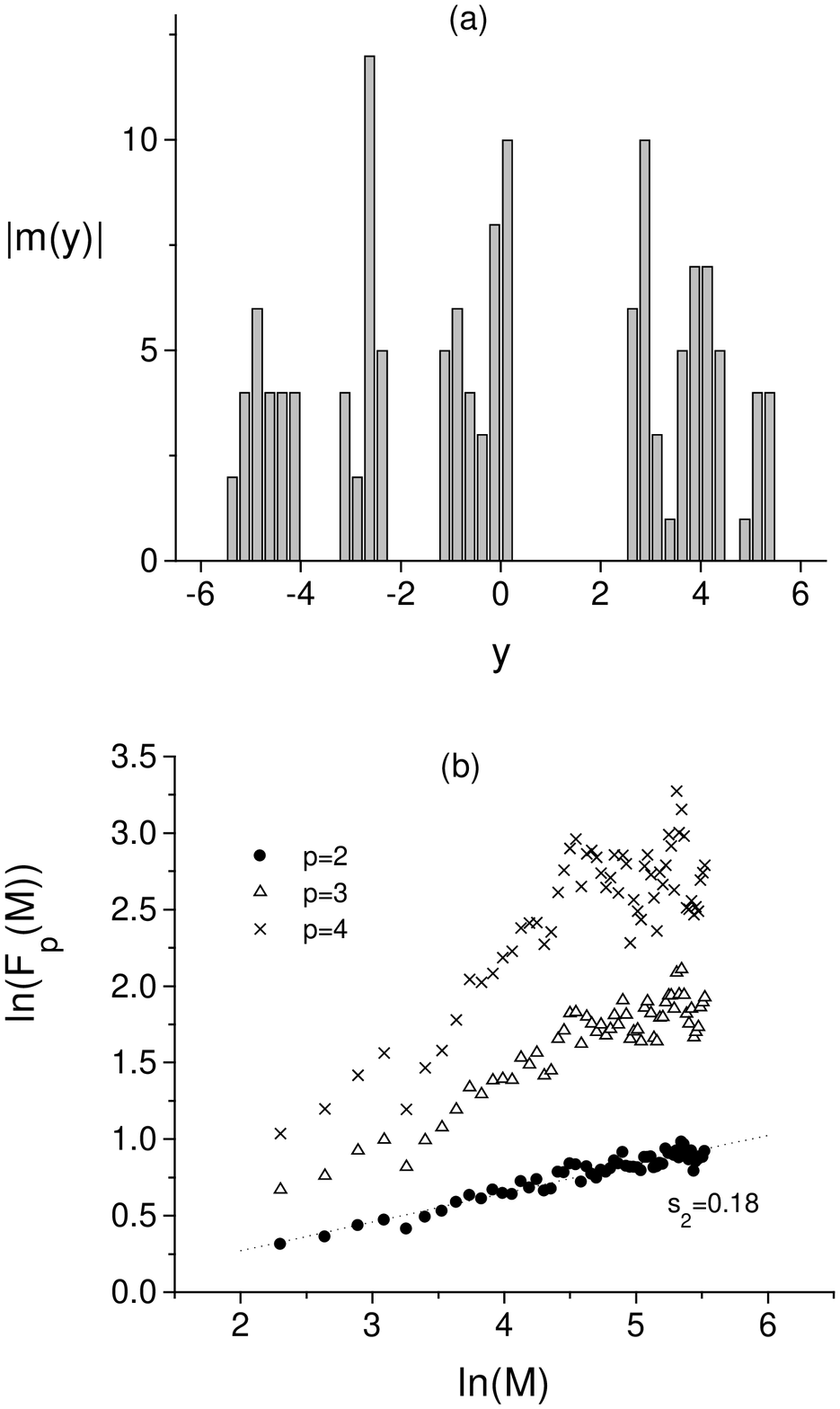,width=0.9\textwidth,angle=0}}
\caption{(a) The net-baryon fluctuations in rapidity space for a typical
critical event. (b) The first three factorial moments for the net-baryon
distribution in rapidity corresponding to the event shown in (a).}
\label{fig:2ab}
\end{center}
\end{figure}

In Fig.~3 the inclusive distribution of net baryons  is illustrated,
built-up by $10^3$ underlying critical events. As expected, an extended plateau is 
generated in rapidity with height $n_b \approx 15$, in accordance with the diagram in 
Fig.~1.

\begin{figure}[htb]
\begin{center}
\mbox{\epsfig{file=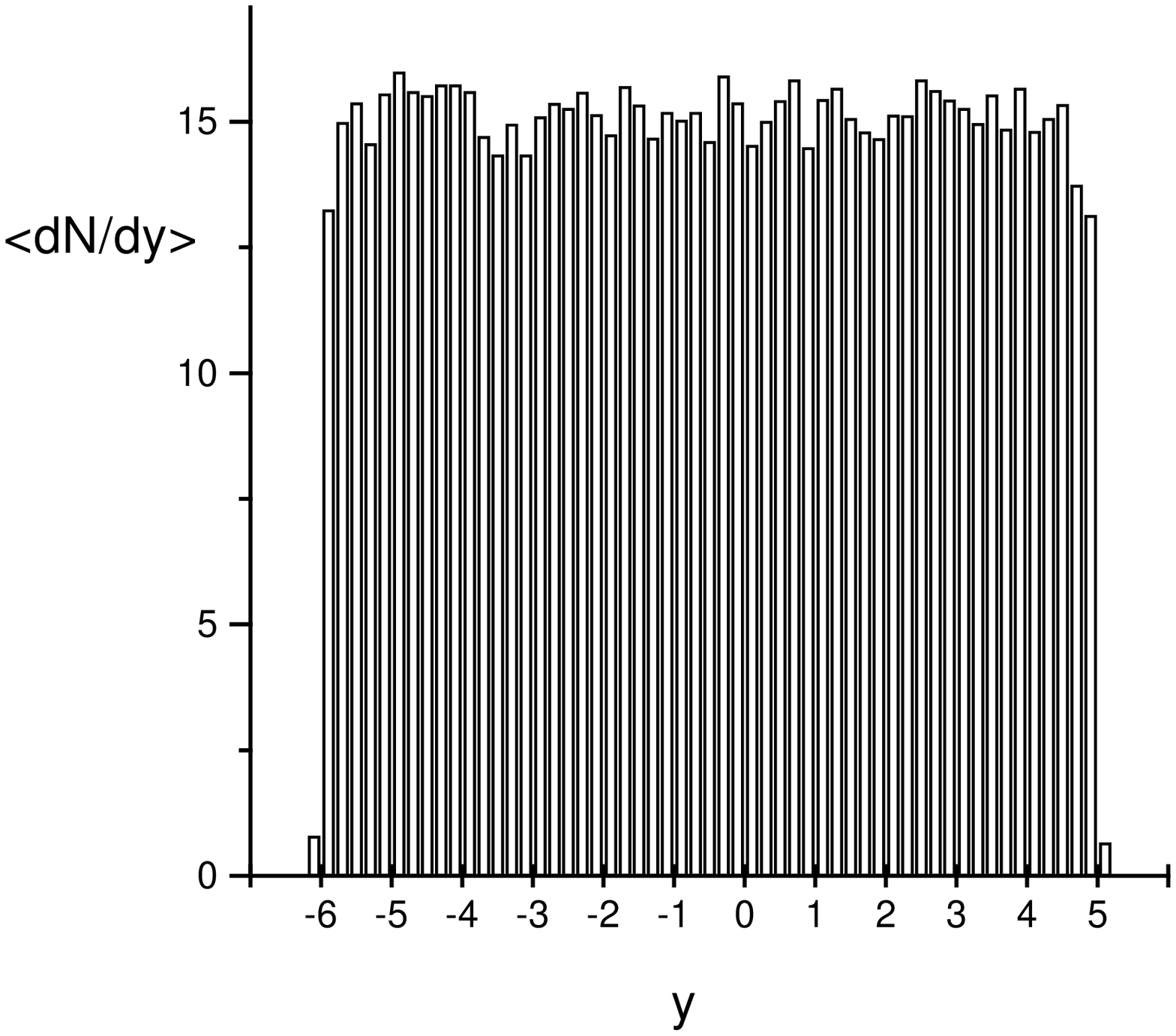,width=0.9\textwidth,angle=0}}
\caption{The inclusive net-baryon distribution in rapidity for 1000 MC
generated critical events. The parameters in the MC have been chosen in
accordance with the conditions at RHIC.}
\label{fig:3}
\end{center}
\end{figure}

In conclusion, we have argued that a sharp signature of the QCD critical point at RHIC,
consists of the following phenomena in the net-baryon sector:

\begin{itemize}
\item{An extended plateau of net baryons in rapidity with height $n_b \approx 15$}
\item{Strong intermittency fluctuations of net baryons in rapidity with index 
$s_2 \approx 1/6$}
\end{itemize}

The absence of these phenomena at RHIC ($Au+Au$, $\sqrt{s}=200~GeV$) would indicate that, in
these experiments, quark matter approaches the critical point in a state far from thermal
equilibrium.


\section*{References}
\begin{thebibliography}{99}

\bibitem{Wil00} F. Wilczek, hep-ph/0003183; J. Berges and K. Rajagopal,
\Journal{\NPB}{538}{215}{1999}; M.A. Stephanov, K. Rajagopal and E. Shuryak, 
\Journal{\PRL}{81}{4816}{1998}.

\bibitem{Fod01} Z. Fodor and S.D. Katz, hep-lat/0106002.

\bibitem{Ant01} N.G. Antoniou, Y.F. Contoyiannis, F.K. Diakonos, A.I. Karanikas
and C.N. Ktorides, \Journal{\NPA}{693}{799}{2001}.

\bibitem{Anto01} N.G. Antoniou, \Journal{\NPB}{92}{26}{2001}.

\bibitem{Ant98} N.G. Antoniou, Y.F. Contoyiannis, F.K. Diakonos and
C.G. Papadopoulos, \Journal{\PRL}{81}{4289}{1998}; N.G. Antoniou, 
Y.F. Contoyiannis and F.K. Diakonos, \Journal{\PRE}{62}{3125}{2000}.

\bibitem{Bia86} A. Bialas and R. Peschanski, \Journal{\NPB}{273}{703}{1986};
\Journal{\NPB}{308}{857}{1988}; E.A. De Wolf, I.M. Dremin and W. Kittel, 
\Journal{\PRP}{270}{1}{1996}.

\end{thebibliography}
\end{document}